# Redshift Cosmológico en un Modelo Bianchi I Axisimétrico: Análisis Cualitativo

**López, Ericson [1] ; Llerena, Mario[1]**

[1]*Escuela Politécnica Nacional, Unidad de Gravitación y Cosmología, Observatorio Astronómico de Quito*

**Resumen:** Dado que la evidencia de las anomalías de la Radiación Cósmica de Fondo sugiere que el Universo es ligeramente anisótropo, el planteamiento de un Modelo Cosmológico alternativo al modelo que cumple el Principio Cosmológico es necesario. Los modelos Bianchi anisótropos se han estudiado puesto que serían necesarios para entender las etapas iniciales del Universo. En este trabajo analizamos cualitativamente el comportamiento del redshift cosmológico en un modelo Bianchi tipo I tras un cambio de coordenadas en tres casos: el vacío con Λ≠0, época de dominio de polvo y época de dominio de radiación.

**Palabras clave:** Cosmología, Anisotropía, Redshift Cosmológico, Bianchi I.

# Cosmological Redshift in a Bianchi I Axisymmetric Model: A Qualitative Analysis

**Abstract:** Cosmic Microwave Background anomalies suggest that the Universe is slightly anisotropic, so a Cosmological Model that does not satisfy the Cosmological Principle is necessary. Anisotropic Bianchi models have been studied given that they would be necessary to understand the early stages of the Universe. In this paper, we analyze qualitatively the behavior of the cosmological redshift in a Bianchi type I model after a coordinate transformation in three cases: vacuum with no vanishing cosmological constant Λ≠0, universe dominated by dust and universe dominated by radiation.

**Keywords:** Cosmology, Anisotropy, Cosmological Redshift, Bianchi I.

## 1. INTRODUCCIÓN

El Modelo Cosmológico Estándar se construye a partir del Principio Cosmológico, según el cual, el Universo es isótropo y homogéneo. Pero los resultados obtenidos por las misiones COBE, WMAP y Planck sobre el estudio de la Radiación Cósmica de Fondo (CMB, por sus siglas en inglés) sugieren que el Universo presenta ligeras anisotropías que no son explicadas a partir del modelo isótropo y homogéneo basado en la métrica de Friedmann-Lemaitre-Robertson-Walker (FLRW) (Akarsu et al., 2010). Para hallar soluciones a este problema abierto de la Cosmología, se plantean universos más generales que relajen sus condiciones de simetría e incluyan la anisotropía como una característica intrínseca de su geometría espacial.

Ante este escenario de incompatibilidad entre el modelo y las observaciones, los modelos Bianchi, los cuales son modelos cosmológicos homogéneos y anisótropos, se han planteado como alternativas para resolver estas anomalías. Además, la comprensión de las anisotropías permite describir el mecanismo que produjo el Big Bang y explicar las ligeras fluctuaciones de temperatura de la CMB (Menezes et al., 2013).

Particularmente, los modelos Bianchi tipo I son analizados porque contienen al Modelo Estándar FLRW con geometría espacial plana (Russell et al., 2014; Saha, 2005) y por su gran importancia para la descripción de las etapas iniciales del Universo (Chawla et al., 2013). Un resultado que se ha hallado con los modelos Bianchi tipo I es que la evolución del universo tiende a isotropizarlo a pesar de la presencia de anisotropías en etapas tempranas (Pradhan et al., 2015).

La anisotropía en este tipo de modelos es tratada al introducir varios factores de escala temporales en la métrica. Cabe recordar que en el Modelo Cosmológico Estándar se tiene un solo factor de escala que determina la evolución temporal del universo. De esta forma, en coordenadas cartesianas y como se muestra en (Kandalkar et al., 2009; Pradhan et al., 2015), el modelo Bianchi I tienen una métrica de la forma que se muestra en la Ecuación (1)

$$ds^2 = dt \otimes dt - a^2(t)dx \otimes dx - b^2(t)dy \otimes dy - c^2(t)dz \otimes dz \quad (1)$$

donde a(t), b(t), c(t) son factores de escala que no necesariamente son iguales. Se puede notar que se introduce un factor diferente en cada coordenada cartesiana, lo que permite que la evolución en las escalas en los ejes cartesianos sea independiente entre sí y por lo tanto, espacialmente, se pierda la simetría esférica, lo que se espera en un modelo anisótropo en el cual las direcciones no son equivalentes entre sí, al menos, en principio.

Como se muestra en (López et al., 2016), a partir de una transformación de coordenadas del tipo indicado en la Ecuación (2),

$$x = r'a(t)cos\theta sin\phi$$
$$y = r'b(t)sin\theta sin\phi \quad (2)$$
$$z = r'c(t)cos\phi$$

con r' ≥ 0 la distancia comóvil, $0 \leq \theta \leq 2\pi$ y $0 \leq \varphi \leq \pi$, y si se considera el caso axisimétrico a(t) = b(t), la métrica en la Ecuación (1), en las nuevas coordenadas, es la indicada en el Ecuación (3):

$$ds^2 = dt \otimes dt - (a^2\sin^2\phi + c^2\cos^2\phi)dr' \otimes dr'$$
$$- r'^2 a^2 \sin^2\phi \, d\theta \otimes d\theta$$
$$- r'^2(c^2\sin^2\phi + a^2\cos^2\phi)d\phi \otimes d\phi$$
$$- 2r'(a^2 - c^2)\sin\phi\cos\phi \, dr' \otimes d\phi \quad (3)$$

Para este caso, considerando un fluido ideal en reposo como fuente de campo gravitatorio con presión p y densidad de materia-energía $\rho$ y, además, con constante cosmológica $\Lambda$ no nula, las ecuaciones de campo son las indicadas en las Ecuaciones (4), (5) y (6):

$$\frac{a^2c\Lambda + c\dot{a}^2 + 2a\dot{a}\dot{c}}{a^2c} = k\rho \quad (4)$$

$$\frac{c^3\dot{a}^2 - a^3\dot{a}\dot{c} - a^4\ddot{c} - (a^4c - a^2c^3)\Lambda - (a^3c - 2ac^3)\ddot{a}}{a^2c} = kp(a^2 - c^2) \quad (5)$$

$$\frac{a^2c\Lambda + ac\ddot{a} + a\dot{a}\dot{c} + a^2\ddot{c}}{c} = -kpa^2 \quad (6)$$

que son las correspondientes a las ecuaciones de Friedmann en el caso axisimétrico que se plantea. $k$ es una constante que se obtiene del límite newtoniano.

Además, de la ley de conservación $\nabla_\mu T^{\mu\nu} = 0$, donde $T^{\mu\nu}$ es el tensor energía-momento del fluido ideal, en la Ecuación (7) se tiene que

$$\dot{\rho} + (p + \rho)\left(2\frac{\dot{a}}{a} + \frac{\dot{c}}{c}\right) = 0 \quad (7)$$

Se considera un fluido ideal en reposo como fuente de materia pues estamos interesados en conocer el comportamiento del redshift cosmológico en un universo anisótropo y no en el efecto de las fuentes anisótropas sobre este parámetro, es decir, estamos interesados en conocer si en un universo anisótropo con fuentes de materia isótropas el redshift cosmológico varía su comportamiento respecto al universo isótropo con la misma fuente materia.

En este trabajo se realiza un análisis cualitativo del comportamiento del redshift cosmológico en un universo Bianchi tipo I axisimétrico empleando la métrica en la Ecuación (3). Se estudian tres casos particulares: vacío con constante cosmológica no nula, época de dominio de polvo y época de dominio de radiación, estos dos últimos casos con constante cosmológica nula.

## 2. METODOLOGÍA

En esta sección se determina una expresión general para el redshift cosmológico en un universo Bianchi I a partir de la métrica en la Ecuación (3). A partir de esta expresión se realiza el análisis cualitativo en casos particulares.

### 2.1 Redshift Cosmológico Anisótropo

Dado que la métrica en la Ecuación (3) se encuentra en coordenadas esféricas, es posible hacer cortes angulares en la variedad.

Considerando superficies con $\varphi = \varphi_0$ y $\theta = \theta_0$ ($\varphi_0$ y $\theta_0$ constantes) en la Ecuación (8) se tiene la métrica inducida en dichas superficies:

$$ds^2_{\phi_0,\theta_0} = dt \otimes dt - (a^2\sin^2\phi_0 + c^2\cos^2\phi_0)dr' \otimes dr' \quad (8)$$

En este caso, considerando el movimiento radial de un haz de luz emitido en un tiempo $t_1$ en la posición r' = R y observado en un tiempo $t_0$ en la posición r' = 0, como se muestra en la Ecuación (9), se tiene que

$$\int_{t_1}^{t_0} \frac{dt}{\sqrt{a^2\sin^2\phi_0 + c^2\cos^2\phi_0}} = \int_R^0 dr' \quad (9)$$

Por otro lado, como se indica en la Ecuación (10), la siguiente cresta de la onda asociada a la radiación llegará a la posición r' = R en un tiempo $t_1 + \lambda_1$, donde $\lambda_1$ es la longitud de onda durante la emisión y mientras que será observada en la posición r' = 0 en un tiempo $t_0 + \lambda_0$ donde $\lambda_0$ es la longitud de onda que se observa, es decir,

$$\int_{t_1+\lambda_1}^{t_0+\lambda_0} \frac{dt}{\sqrt{a^2\sin^2\phi_0 + c^2\cos^2\phi_0}} = \int_R^0 dr' \quad (10)$$

Por lo tanto, igualando las Ecuaciones (9) y (10) se tiene el resultado mostrado en la Ecuación (11):

$$\int_{t_1}^{t_0} \frac{dt}{\sqrt{a^2\sin^2\phi_0 + c^2\cos^2\phi_0}} = \int_{t_1+\lambda_1}^{t_0+\lambda_0} \frac{dt}{\sqrt{a^2\sin^2\phi_0 + c^2\cos^2\phi_0}} \quad (11)$$

Después de un cambio en los límites de integración en la Ecuación (11) se tiene la Ecuación (12) que se muestra a continuación:

$$\int_{t_0}^{t_0+\lambda_0} \frac{dt}{\sqrt{a_0^2\sin^2\phi_0 + c_0^2\cos^2\phi_0}} = \int_{t_1}^{t_1+\lambda_1} \frac{dt}{\sqrt{a_1^2\sin^2\phi_0 + c_1^2\cos^2\phi_0}} \quad (12)$$

donde los subíndices 0 y 1 en los factores de escala representan el valor de los mismos en el tiempo de observación y en el tiempo de emisión, respectivamente. Considerando que $a_0$ y $c_0$ son constantes entre $t_0$ y $t_0 + \lambda_0$, y $a_1$ y $c_1$ lo son entre $t_1$ y $t_1 + \lambda_1$, entonces a partir de la Ecuación (12) se tiene la Ecuación (13)

$$\frac{\lambda_0}{\sqrt{a_0^2\sin^2\phi_0 + c_0^2\cos^2\phi_0}} = \frac{\lambda_1}{\sqrt{a_1^2\sin^2\phi_0 + c_1^2\cos^2\phi_0}} \quad (13)$$

Con esto, dado que se conoce la relación expuesta en la Ecuación (14):

$$z + 1 = \frac{\lambda_0}{\lambda_1} \quad (14)$$

se tiene que el redshift cosmológico en el modelo propuesto está dado por la Ecuación (15)

$$1 + z = \frac{\sqrt{a_0^2\sin^2\phi_0 + c_0^2\cos^2\phi_0}}{\sqrt{a_1^2\sin^2\phi_0 + c_1^2\cos^2\phi_0}} \quad (15)$$

para el universo anisótropo descrito por la Ecuación (3). Por notación, sea el parámetro definido en la Ecuación (16):

$$A_0 = \sqrt{a_0^2\sin^2\phi_0 + c_0^2\cos^2\phi_0} \quad (16)$$

que depende de los valores actuales de los factores de escala, se tiene que el redshift cosmológico está dado por la Ecuación (17):

$$\frac{1+z}{A_0} = \frac{1}{\sqrt{a_1^2\sin^2\phi_0 + c_1^2\cos^2\phi_0}} \quad (17)$$

Se puede notar que el redshift cosmológico tiene dependencia angular con respecto al ángulo φ, lo cual no ocurre en el modelo isótropo, en donde se tiene que $1 + z = a_0/a_1$. En la siguiente sección se usa la Ecuación (17) del redshift cosmológico normalizado respecto a los valores actuales de $A_0$ para describir su evolución en casos particulares.

## 3. RESULTADOS Y DISCUSIÓN

En esta sección se realiza un análisis cualitativo del redshift cosmológico a partir de la Ecuación (17). Se analizan tres casos particulares para poder discutir las diferencias halladas con el comportamiento de acuerdo al modelo isótropo y homogéneo.

### 3.1. *Universo Vacío Con Constante Cosmológica*

En el vacío se tiene que $T_{\mu\nu} = 0$. Como se muestra en (López et al., 2016), las ecuaciones de campo en el vacío con Λ≠0 se resuelven cuando los factores de escala son los mostrados en la Ecuación (18):

$$a(t) = K_a e^{\alpha t} \quad c(t) = K_c e^{\alpha t} \quad (18)$$

con Ka y Kc constantes. Además, se cumple que el parámetro $\alpha = \sqrt{-\Lambda/3}$ corresponde al parámetro de Hubble (para la solución de vacío).

Para este análisis cualitativo se considera el parámetro de Hubble como α = 1, muy diferente al valor que tendría si tomamos el valor que se ha estimado de $\Lambda = 10^{-122}$ en unidades naturales para la constante cosmológica (Barrow et al, 2011), esto debido a que estamos interesados en un análisis cualitativo por el momento. Por otro lado, se considera Ka = 1 tomando la misma normalización que se acostumbra para el modelo estándar, es decir, si t=0 entonces a=1. En la Figura 1, se muestra la evolución temporal del redshift cosmológico en un universo Bianchi I axisimétrico vacío con Λ≠0 para distintos valores de φ₀ y Kc = 0.5 fijo.

Se puede notar de la Ecuación (17) que si Ka = Kc, la evolución temporal no depende del ángulo φ₀ pero, en el caso axisimétrico, donde Kc no es necesariamente igual a Ka, se puede evidenciar que la evolución es diferente para cada plano φ = φ₀, siendo simétrica para φ = −φ₀.

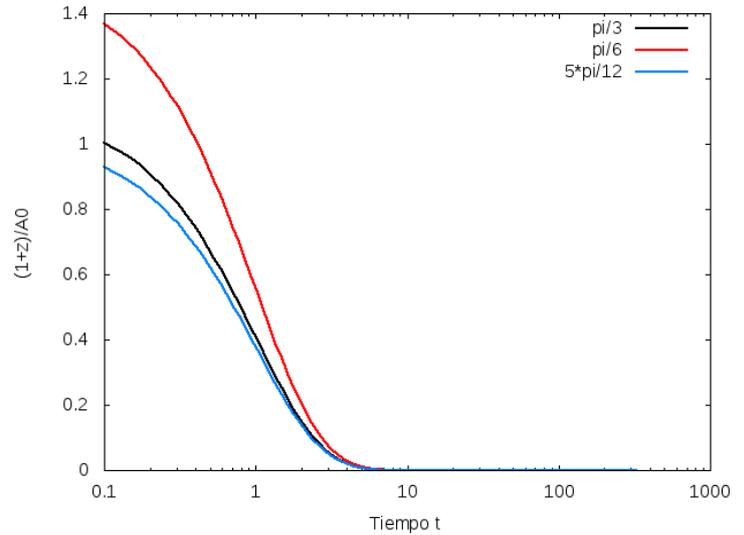

**Figura 1**. Evolución temporal del redshift cosmológico para la solución de vacío con α= 1, Ka = 1, Kc = 0.5 y distintos valores para φ₀. Para t=0, se tiene un valor finito.

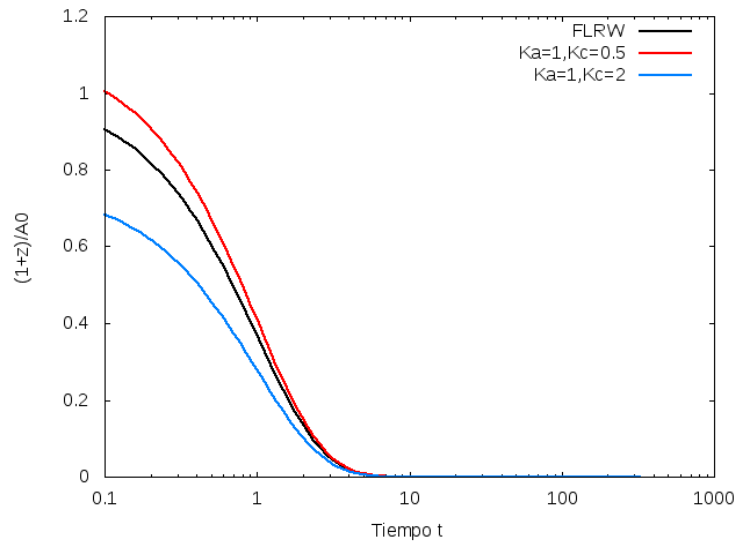

**Figura 2.** Evolución temporal del redshift cosmológico para la solución de vacío con α= 1, Ka = 1, φ₀=π/3 y distintos valores para Kc. Para t=0, se tiene un valor finito.

En la Figura 2, se muestra dicha evolución, pero para distintos valores de Kc y φ₀ = π/3 fijo. Con línea negra continua se señala el caso isótropo y homogéneo FLRW. En este caso se encuentra que el redshift cosmológico es mayor que en el caso FLRW (Ka=Kc=1) siempre que Kc < 1 cuando Ka = 1, al menos en etapas iniciales del universo. Si Kc > 1 cuando Ka = 1, el redshift cosmológico es menor que en el universo isótropo, del mismo modo, al menos en etapas tempranas. Cuanto t=0, se tiene un valor finito diferente para cada caso.

En ambos casos, tanto variando $\varphi_0$ como Kc, se tiene que mientras evoluciona el universo, el redshift cosmológico tiende a un mismo valor en los distintos casos, es decir, tiende al mismo comportamiento del universo isótropo. Además, si $\Lambda = 0$, el redshift toma un valor constante en cada caso y no hay proceso de isotropización.

Con estos resultados se hace evidente que en un universo anisótropo vacío, la caracterización del redshift cosmológico permite establecer la constante de proporcionalidad entre los factores de escala y además, es posible enunciar que la constante cosmológica es fundamental en la evolución del universo vacío.

### 3.2. *Época De Dominio Del Polvo*

Si se considera que el universo está dominado por polvo, es decir, partículas no interactuantes, se tiene que p=0 y con esta ecuación de estado es posible plantear las ecuaciones de campo correspondiente para esta etapa del universo dominada por este fluido. Considerando $\Lambda = 0$ (a pesar de contar con un valor estimado para la constante cosmológica (Barrow et al, 2011)), es decir, analizando únicamente la contribución del polvo mas no de otras fuentes en la evolución del universo, como se muestra en (López et al., 2016), se puede verificar que una propuesta del tipo que se indica en la Ecuación (19)

$$a(t) = K_a t^{2/3} \qquad c(t) = K_c t^{2/3} \qquad (19)$$

donde Ka y Kc son constantes, satisface las ecuaciones de campo y, adicionalmente, por la ley de conservación, se tiene que la densidad de materia-energía decae como en la Ecuación (20):

$$\rho(t) = \frac{4}{3\kappa} t^{-2} \qquad (20)$$

Considerando Ka = 1 (sólo por motivos de comportamiento cualitativo), se puede analizar la evolución de redshift cosmológico al variar $\varphi_0$ y Kc en la época de dominio de polvo sin constante cosmológica.
En la Figura 3, se muestra dicha evolución con Kc = 0.5 y para distintos valores de $\varphi_0$, mientras que en la Figura 4, se muestra la evolución con $\varphi_0 = \pi/3$ y para distintos valores de Kc. Cuando t=0, se tiene un valor no finito.
Se puede notar que, en etapas iniciales, a diferencia del universo vacío con constante cosmológica, el comportamiento del modelo anisótropo es similar al modelo FLRW que se señala con línea negra continua y lo mismo ocurre en etapas posteriores de su evolución, pero existen etapas en las cuales los modelos son diferentes.

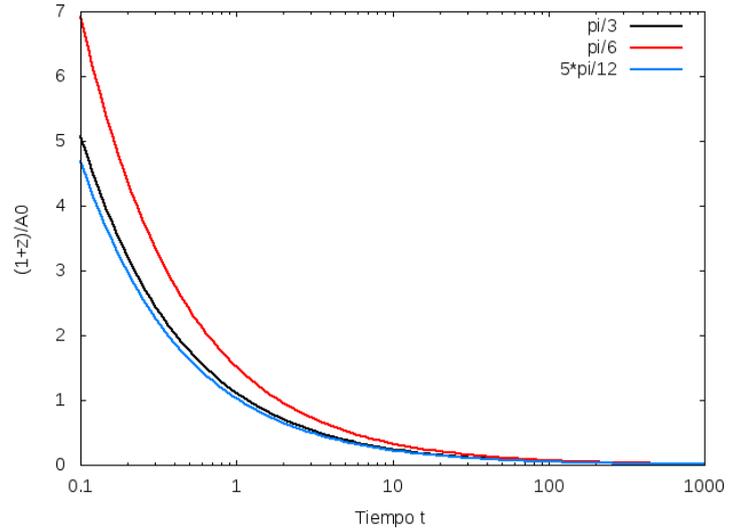

**Figura 3.** Evolución temporal del redshift cosmológico para la solución de dominio de polvo con Ka = 1, Ka = 0.5 y distintos valores para $\varphi_0$. Para t=0, se tiene un valor no finito.

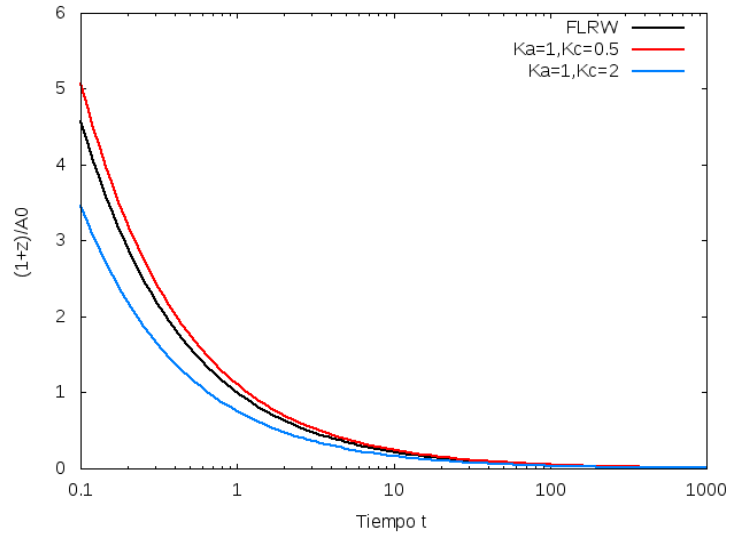

**Figura 4.** Evolución temporal del redshift cosmológico para la solución de dominio de polvo con Ka = 1, $\varphi_0 = \pi/3$ y distintos valores para Kc. Para t=0, se tiene un valor no finito.

Es decir, el universo dominado por polvo inicialmente es isótropo pues el redshift cosmológico tiene igual valor que en el modelo FLRW, pero en alguna etapa de su evolución se vuelve ligeramente anisótropo para posteriormente volverse isótropo nuevamente. Así mismo, se determina la dependencia angular en el redshift cosmológico, lo que no ocurre en el modelo isótropo.

### 3.3. *Época Dominada Por Radiación*

Si se considera que el universo está dominado por radiación que cumple con la ecuación de estado p = $\rho/3$ y considerando que $\Lambda = 0$, como en (López et al., 2016), se puede verificar que una solución del tipo mostrado en la Ecuación (21):

$$a(t) = K_a t^{1/2} \qquad c(t) = K_c t^{1/2} \qquad (21)$$

resuelve las ecuaciones de campo para esta etapa del universo de dominio de radiación. Adicionalmente, por la ley de continuidad, se cumple que la densidad de materia-energía evoluciona temporalmente como en la Ecuación (22):

$$\rho(t) = \frac{3}{4\kappa} t^{-2} \quad (22)$$

Considerando Ka = 1 para un análisis cualitativo, se puede analizar la evolución de redshift cosmológico al variar $\varphi_0$ y Kc. En la Figura 5, se muestra dicha evolución con Kc = 0.5 y para distintos valores de $\varphi_0$, mientras que en la Figura 6, se muestra la evolución con $\varphi_0 = \pi/3$ y para distintos valores de Kc.

Se puede notar que, en este caso, cuando t=0, se tiene un valor no finito y se repiten las etapas de pérdida de isotropía y posterior isotropización, tal como se describió en la etapa de dominio de polvo.

Es decir, las fuentes de campo gravitatorio de fluido ideal permiten que la evolución del redshift cosmológico sea la misma, tanto en un universo isótropo como en uno anisótropo, al menos en etapas iniciales, lo cual no ocurre en el universo vacío. Por otro lado, hay diferencias en su comportamiento, respecto al universo isótropo, en alguna etapa de su evolución.

En los casos con fuentes de materia se puede pensar que la isotropía inicial del universo depende de la nulidad de la constante cosmológica, comparando con los resultados del universo vacío con constante cosmológica.

Por otro lado, en el caso de dominio de radiación, la etapa de isotropización ocurre en un tiempo mayor comparado con la etapa de dominio de polvo, es decir, el redshift se isotropiza en un menor tiempo en un universo dominado por polvo que en uno dominado por radiación.

En los casos de dominio de polvo y radiación se puede verificar que se repite el comportamiento temporal del redshift cosmológico, en referencia a su valor dependiente del plano $\varphi_0$ en el caso anisótropo y además, su valor es, en alguna etapa de su evolución, mayor que en el universo FLRW si el universo es alargado en el plano XY (Ka>Kc) y es menor para el caso donde el universo es alargado con respecto al eje de simetría (Ka<Kc).

Estos resultados cualitativos del comportamiento temporal del redshift cosmológico son importantes dado que tienen implicaciones en la medición de distancias cosmológicas, la cuales son funciones de este parámetro (Hogg, 1999). Una subestimación o una sobrestimación de su valor tiene consecuencias en el valor de la distancia calculada para objetos lejanos, así como para sus velocidades, especialmente en las etapas de no isotropía. En una próxima contribución se plantearán las implicaciones del modelo anisótropo en las distancias cosmológicas.

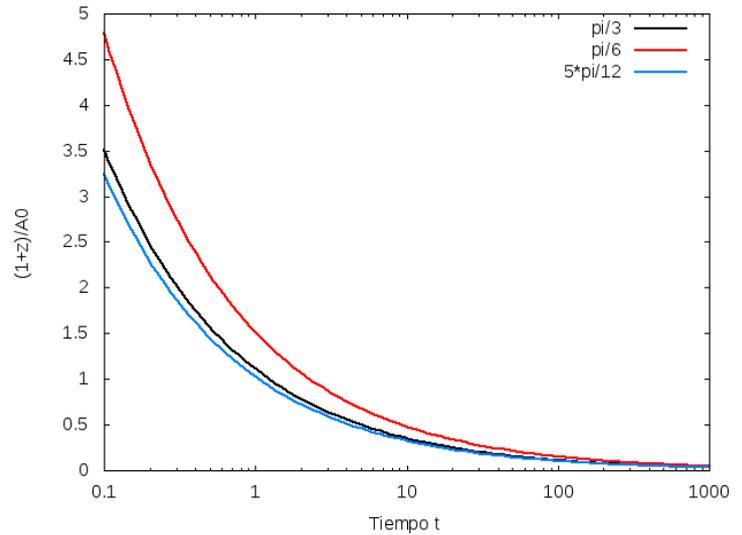

**Figura 5.** Evolución temporal del redshift cosmológico para la solución de dominio de radiación con Ka = 1, Ka = 0.5 y distintos valores para $\varphi_0$. Para t=0, se tiene un valor no finito.

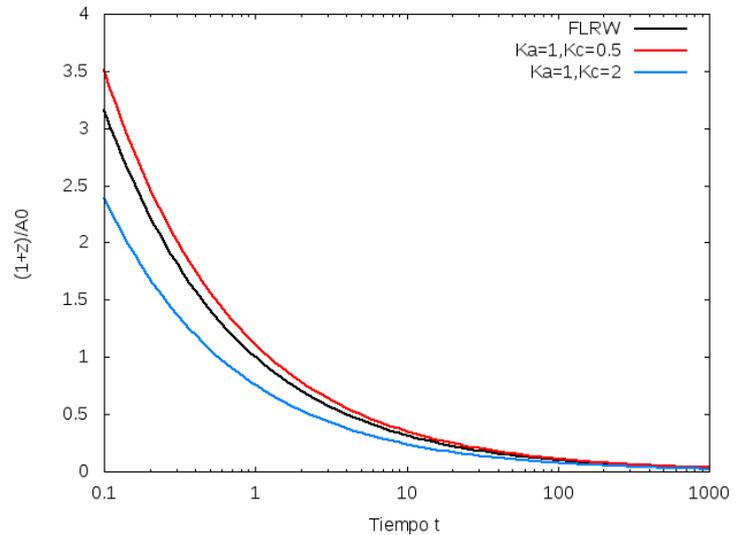

**Figura 6.** Evolución temporal del redshift cosmológico para la solución de dominio de radiación con Ka = 1, $\varphi_0 = \pi/3$ y distintos valores para Kc. Para t=0, se tiene un valor no finito.

## 4. CONCLUSIONES

El redshift cosmológico en un universo Bianchi I axisimétrico se ha analizado cualitativamente. Para la determinación del redshift cosmológico se empleó una métrica en coordenadas esféricas reducible al caso FLRW isótropo con geometría espacial plana. Se consideró el caso axisimétrico con dos factores de escala.

Se determinó que la evolución temporal del redshift cosmológico depende del ángulo $\varphi_0$ de la superficie de la variedad, y dicha dependencia es simétrica con respecto al plano XY.

Se determinó que en el proceso de evolución del redshift se presentan etapas de isotropización. En el caso del universo vacío con constante cosmológica, se estableció que en etapas iniciales los valores del redshift cosmológico difiere en los casos anisótropos e isótropos, pero en tiempos posteriores tienden a un valor cercano.

En los casos de dominio de polvo y radiación se encontró el mismo comportamiento para tiempos posteriores en su evolución, pero en etapas iniciales se halló que el comportamiento entre un universo isótropo y uno anisótropo es el mismo. Es decir, hay una etapa de pérdida de isotropía y luego isotropización en el redshift en un universo Bianchi I axisimétrico dominado por polvo o radiación y con constante cosmológica nula.

En cuanto a la constante cosmológica se puede concluir que, en el vacío, contribuye para que el universo sea anisótropo inicialmente, mas no ocurre lo mismo con el aporte del contenido de fluido ideal barotrópico (polvo y radiación).

Finalmente, es importante mencionar que, como recomendación, es necesario hacer un análisis cuantitativo del redshift cosmológico pues esto permitiría determinar en que fase de la evolución del universo nos encontramos actualmente. Así mismo, es importante determinar si actualmente estamos en una etapa isótropa o anisótropa.

El redshift cosmológico podría ser una magnitud física que permita evidenciar anisotropías.

## REFERENCIAS

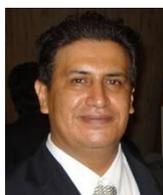
**Ericson López.** Doctor en Astrofísica (PhD) por la Academia de Ciencias de Rusia y Físico Teórico por la Escuela Politécnica Nacional. Ha realizado investigaciones post doctorales en Brasil y Estados Unidos. Es científico colaborador del Harvard-Smithsonian Center para la Astrofísica y profesor adjunto del Departamento de Astronomía de la Universidad de Sao Paulo. Ha realizado más de 30 publicaciones científicas y varias otras publicaciones relevantes. Es Director del Observatorio Astronómico de Quito desde 1997 y miembro de la Academia de Ciencias del Ecuador. Es profesor principal de la Facultad de Ciencias de la EPN por más de 25 años, en la que imparte cursos formales de Física y Astrofísica.

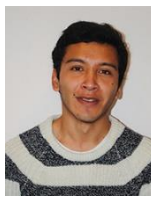
**Mario Llerena.** Realizó sus estudios de pregrado en la Escuela Politécnica Nacional, obteniendo su título de Físico en el 2016. Es miembro del Observatorio Astronómico de Quito como parte de Unidades Científicas de investigación en Gravitación y Cosmología, Radioastronomía y Astrofísica de Altas Energías.